\documentclass[a4paper,UKenglish,cleveref, autoref, thm-restate]{lipics-v2021}

\usepackage{hyperref}
\usepackage{float}
\nolinenumbers
\title{Modern Constraint Programming Education: Lessons for the Future} 

\author{Tejas {Santanam}}{Georgia Institute of Technology, Atlanta, GA, USA \and \url{https://sites.gatech.edu/tsantanam/}} {tsantanam@gatech.edu}{https://orcid.org/0000-0002-0269-6510}{}

\author{Pascal {Van Hentenryck}}{Georgia Institute of Technology, Atlanta, GA, USA \and \url{https://sites.gatech.edu/pascal-van-hentenryck/}} {pvh@gatech.edu}{https://orcid.org/0000-0001-7085-9994}{}

\authorrunning{T. Santanam and P. Van Hentenryck} 

\Copyright{Tejas Santanam and Pascal van Hentenryck} 



\category{} 

\relatedversion{} 

\acknowledgements{Thanks to Yoda for inspiring many and Helmut Simonis for co-organizing this workshop. And of course, none of this would be possible without the many students we have taught over the years. May the Force be with them.}

\EventEditors{Tejas Santanam and Helmut Simonis}
\EventNoEds{2}
\EventLongTitle{Workshop on Teaching Constraint Programming, WTCP 2023}
\EventShortTitle{WTCP 2023}
\EventAcronym{WTCP}
\EventYear{2023}
\EventDate{August 27th, 2023}
\EventLocation{Toronto, Canada}
\EventLogo{}
\SeriesVolume{1}
\ArticleNo{1}

\begin{document}

\maketitle

\begin{abstract}
This paper details an outlook on modern constraint programming (CP) education through the lens of a CP instructor. A general overview of current CP courses and instructional methods is presented, with a focus on online and virtually-delivered courses. This is followed by a discussion of the novel approach taken to introductory CP education for engineering students at large scale at the Georgia Institute of Technology (Georgia Tech) in Atlanta, GA, USA. The paper summarizes important takeaways from the Georgia Tech CP course and ends with a discussion on the future of CP education. Some ideas for instructional methods, promotional methods, and organizational changes are proposed to aid in the long-term growth of CP education.
\end{abstract}

\section{Introduction and Current CP Education}
\label{sec:introduction}

Constraint programming (CP) is a methodology for solving combinatorial problems to find feasible or optimal solutions by using constraints to reduce the set of values each variable in a problem can potentially take. The field lies at the intersection of operations research and computer science and drives numerous applications in the real world, such as hospital scheduling, sports tournament bracketing, delivery vehicle routing, and evacuation planning.

In general, CP education at the university level tends to be fairly decentralized. Unlike the abundance of machine learning courses, for example, that one can find at almost every university today, existence of CP courses is largely predicated on having qualified and motivated CP practitioners willing to design or teach such a course. For many years, quality CP education was largely limited to graduate students who were fortunate enough to work in departments that offered CP courses. Christine Solnon led an online artificial intelligence (AI) course for graduate students in France in the early 2000s that contained 12 hours worth of sessions on Gnu-Prolog and CP. The course focused on the basics of CP modeling and CP solvers while working to solve puzzles like map coloring and ``SEND + MORE = MONEY'' \cite{solnon2004line}. Helmut Simonis started, and still runs, an online course teaching ECLiPSE in order to learn CP modeling and solving techniques. Today, one may self-study using the videos and handouts on the course website \cite{Simonis, simonislessons}. However, in the last decade, the advent of Massive Open Online Courses (MOOCs) brought opportunities to democratize CP education in a structured manner with lessons and assignments for members of the general public. Pascal Van Hentenryck's Discrete Optimization course on Coursera introduced some basics of CP and Large Neighborhood Search (LNS) with over 18 hours of interactive material comprising Indiana Jones-themed videos, readings, and quizzes \cite{Van}. The course also featured the use of an autograder system to grade the vast number of submissions from members of the general public. Johannes Waldmann also developed an `autotool` autograder framework along with some exercises focused on understanding the fundamental algorithms behind CP solvers \cite{waldmann2014automated}. More recently, Jimmy Lee and Peter Stuckey co-developed three Coursera MOOCs on the subject of “Modeling and Solving Discrete Optimization Problems” \cite{chan2020teaching, lee2021mooc}. The courses feature a form of problem-based learning encapsulated in a coherent story plot following classic Chinese novels. Lee and Stuckey presented learner statistics and feedback, and discussed their experience with adopting the online materials in a smaller flipped classroom setting as well. In 2022, Hoffmann et al. produced the first human-centered study addressing how people approach constraint modeling and solving \cite{hoffmann2022understanding}. This information will be useful in pedagogical design for future CP courses. Pierre Schaus, Laurent Michel, and Pascal Van Hentenryck launched a CP MOOC in February 2023 on edX using the Mini-CP solver, a lightweight, open-source CP solver designed for educational purposes \cite{michel2021minicp}. The Mini-CP solver comes with a series of more than 20 implementation projects to help students with the basics of CP modeling and search heuristics.

The effects of the COVID-19 pandemic over the last few years have signaled a shift in the way educational content is delivered and consumed. In order to continue growth as a practice, the CP community must adapt to these new pedagogical styles. This paper draws on the authors' experiences from an online CP course at the Georgia Institute of Technology (Georgia Tech) to detail lessons learned and novel strategies for introductory CP education.

\section{The Georgia Tech CP Course}
\label{sec:gtcp}

The Georgia Tech CP Course was started in Fall 2018 by Pascal Van Hentenryck. The course was initially taught as an in-person course and focused on using the OPL programming language inside the IBM ILOG CPLEX Optimization Studio to model CP problems. The first third of the course focused on learning OPL and modeling puzzles with CP. The second third of the course goes from modeling basic optimization toy problems to condensed versions of real-world optimization problems. The last third of the course focused on CP applications in scheduling, routing, and evacuation planning. The course was housed in the Industrial and Systems Engineering department at Georgia Tech and was open to both advanced undergraduate and graduate students within the department. The course primarily focused on finite-domain CP as an approachable introduction within the College of Engineering. The enrollment of this initial CP course offering was 26 students.

With the COVID-19 pandemic forcing instruction to be delivered online in 2020, Van Hentenryck adapted the method of course delivery considerably in order to maintain effectiveness in a virtual format. This redevelopment modified the Georgia Tech CP course into the form which it takes today. The size of the Georgia Tech CP course grew considerably after this point. While the course was initially offered yearly in the Fall semesters, the frequency was increased to both Fall and Spring semesters starting in Spring 2023. Also in Spring 2023, Tejas Santanam joined the CP course as an instructor, having been a prior student and head teaching assistant for the Georgia Tech CP course.

This is the only CP course offered at any level at Georgia Tech.

The key features of course logistics for the Georgia Tech CP course are detailed below. In total, the course lasts for a full semester (approximately 15 weeks), with an expected 8 to 15 hours of workload each week.

\paragraph*{Course Goals}
The learning outcomes for students are as follows.
\begin{itemize}
    \item Describe the fundamental properties of good constraint programming models and how they differ from other methodologies.
\item Be able to determine when/how to use constraint programming for practical applications in areas such as scheduling, routing, and resource allocation.
\item Achieve fluency in the modeling language OPL for constraint programming.
\item Recognize when additional features (e.g., new constraints and dedicated search procedures) are necessary to solve a problem and understand what this involves.
\end{itemize}

\paragraph*{Topic Outline}
The topics covered through the course are as follows. In some iterations of the course, the advanced topics towards the end of the list were required only for graduate students.
\begin{itemize}
    \item Basic Concepts
    \begin{itemize}
        \item Getting started
\item Basic concepts I
\item Basic concepts II
\item OPL Primer
    \end{itemize}
\item  Elements of Constraint Programming
\begin{itemize}
    \item Reified constraints
\item Optimization
\item Expressions
\end{itemize}
\item  Theoretical Foundation
\begin{itemize}
    \item Computational Model
    \end{itemize}
\item  Global Constraints
\begin{itemize}
    \item The element constraint
\item The table constraint
\item Combinatorial Constraints
\item The pack constraint
\item The circuit constraint
\item The lex constraints
\end{itemize}
\item  Modeling in Constraint Programming
\begin{itemize}
\item Symmetry breaking
\item Subexpression elimination
\item Redundant constraints I
\item Redundant constraints II
\end{itemize}
\item Search in Constraint Programming
\begin{itemize}
    \item Search tree and Impact
\item Restart and nogoods
\end{itemize}
\item Implementation of Constraint Programming
\begin{itemize}
\item Packing
\item AllDifferrent
\item NoOverlap
\end{itemize}
\item Scheduling in Constraint Programming
\begin{itemize}
\item Interval variables and noOverlap
\item The Sequence Constraints
\item Cumulative Constraints
\item The House Problem II
\item The House Problem III
\item The Perfect Square Problem
\item State Constraints
\item The Trolley Application
\item Optional Activities
\item  Standard Scheduling Problems
\item  Calendars
\end{itemize}
\item Advanced Topics
\begin{itemize}
\item  Large neighborhood search
\item  Scripting models
\item  Routing
\item  CP in Python
\end{itemize}
\item Implementation in MiniCP
\begin{itemize}
\item  Semantics of CP
\item  Operational Model of CP
\item  Inference
\item  Search
\item  Advanced Inference
\item  Advanced Search
\end{itemize}
\end{itemize}

\paragraph*{Lecture Videos}
The material in the course is presented in high-quality videos. In total there are about 90 videos comprising around 30 hours of material. Van Hentenryck recorded a video for each topic listed above with greenscreen backgrounds, animations, and more. The videos generally have Van Hentenryck's head superimposed on slides or images, which make for clearer and higher quality videos compared to recordings of traditional classroom proceedings. The videos are relatively short and digestible, with lengths ranging from 10 to 30 minutes each. The students are able to play, pause, and rewind these videos, as well as change the playback speed. A flipped classroom style approach is utilized where the lecture videos are posted in advance of the scheduled class sessions. Lecture videos are complemented with interactive online sessions that happen during the scheduled course time. An example of one of the lecture videos can be seen below in Figure~\ref{fig:video}. The PDF files of the slides used in each lecture video are also provided to the students.

\begin{figure}[H]
\caption{A look at one of the Star Wars-themed lecture videos}
    \centering
        \includegraphics[height=0.6\textwidth]{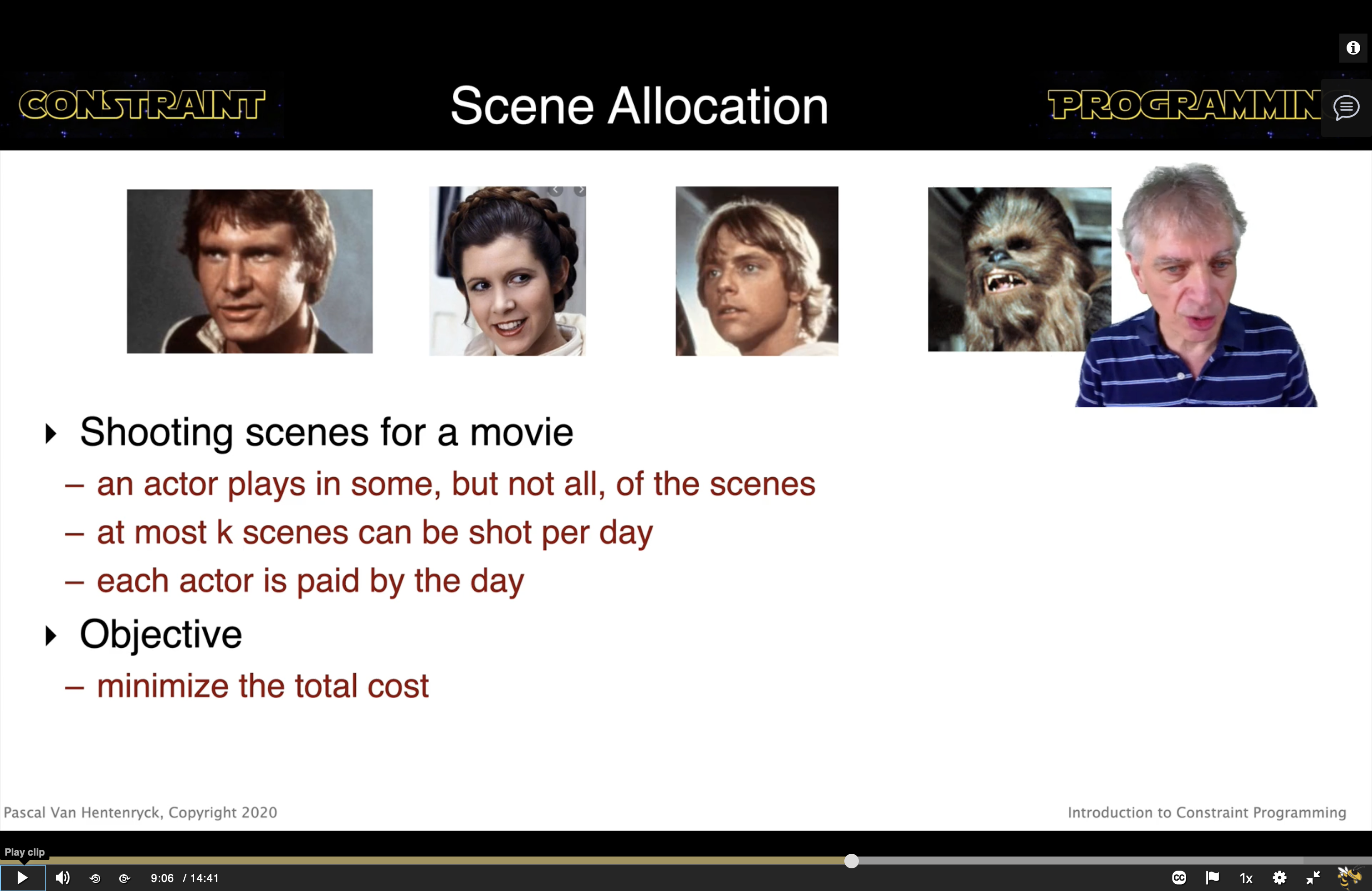}
        \label{fig:video}
        \end{figure}

\paragraph*{Interactive Sessions}
The interactive sessions take place on Zoom and focus on review of the material from the most recent set of lecture videos. The interactive sessions include time for Q\&A and go over the course assignments. They also include one-on-one sessions in Zoom breakout rooms with the instructors and the teaching assistants. These breakout room sessions allow students to get customized face-to-face help. The sessions are frequently used for help with conceptual understanding and debugging code for the assignments. The interactive sessions take place two to three times a week for about an hour each day. Teaching assistants also offer office hours on an ad hoc basis to help those unable to join a particular interactive session.

\paragraph*{Discussion Forums}
The Georgia Tech CP Course also has an attached discussion forum where students can ask conceptual questions publicly and assignment code debugging questions in a private post to course instructors and teaching assistants. Students may answer the questions of their peers or leave follow-up comments and questions to others' questions. Students are guaranteed responses within 24 hours; however, the mean instructor (including teaching assistants) response time is 47 minutes. The response time is often even faster than that during the workday.

\paragraph*{Assignments}
The course contains many assignments that are due on a roughly weekly basis. Each assignment relates to applying the topics covered in the lecture videos and interactive sessions from the previous week. Each assignment can be solved with the knowledge students have learned to that point, as well as with the existing functionalities of OPL and the OPL IDE. Every assignment involves an application-focused modeling problem that requires a model coded in OPL as a solution. The assignments start simple with problems like map coloring before ramping up to more difficult problems like the Capacitated Vehicle Routing Problem with Time Windows (CVRPTW) and flood evacuation planning. Students start with assignments from the very first week to cultivate their constraint programming mindset and familiarize themselves with the OPL language. This is in line with the philosophy espoused by Patrick Prosser in his CP 2014 talk, where he discussed getting students to solve problems as soon as possible \cite{prosser2014teaching}. The focus in the assignments is on modeling and writing appropriate constraints. Students in the course use the solver from IBM ILOG CP Optimizer out of the box, rather than writing custom search procedures. There are a few assignments towards the end of the course that involve CP in Python (due to students' familiarity with Python) and Mini-CP (primarily for graduates students; a way to explore writing search procedures). Some of the introductory assignments are reused year to year, though most of the assignments receive changes ranging from minor tweaks all the way to brand-new exercises. Some of the assignments require data file inputs which are provided to the students. For many of the assignments in the latter half of the course, multiple data files are provided for testing as students devise their formulations. However, students are asked to make their formulation general in nature, as the provided dataset to the students is not necessarily the one used for verifying and grading their model.

\paragraph*{Theme}
Similar to the adoption of an Indiana Jones theme in Van Hentenryck's earlier MOOC \cite{Van} or the classic Chinese novel themes in Lee and Stuckey's MOOCs \cite{chan2020teaching, lee2021mooc}, the Georgia Tech CP course uses a Star Wars theme. All of the lecture videos involve Star Wars-themed examples, graphics, and even costumes! At the time of writing, a viral video of Van Hentenryck teaching while dressed as Yoda has garnered over 1.5 million views on \href{https://www.tiktok.com/@camille_miles/video/6999857222868126981}{TikTok}. The assignments also all have problem descriptions pertaining to happenings in the Star Wars universe. All course instructors are referred to as Jedi Masters (Van Hentenryck is Yoda, while Santanam is Obi-Wan Kenobi), teaching assistants are referred to as Jedi Assistants, and students are referred to as Padawans. At first glance, the theme may seem gimmicky, but it actually serves to motivate the students and keep them engaged with some light-heartedness. There is no requirement of prior Star Wars knowledge, nor does such knowledge provide a leg up in the course. The use of the Star Wars theme, specifically, is allowed for educational use internally. However, it would infringe on Disney's copyright if the lecture videos and assignments were to be public-facing.

\paragraph*{Student Patterns of Interaction}

It is also important to shed light on the manner in which students engage with course material. The graph below in Figure~\ref{fig:interaction} shows the average page views per student per week over the course of a semester. Page views includes actions like viewing a lecture video, reading lecture slides, or going to the interactive session page. The interaction pattern follows a cyclical up and down flow. Students spend more time with the material when it is new, and then focus on the applications as they get more used to the material. Then, when new material is introduced, the page views jump up again. Apart from the drop near the beginning of the semester associated with a school holiday week, interactions until Georgia Tech's final exam period remain consistently above 30 page views per student per week. Page views tend to be higher in the second half of the course when there is a focus on scheduling and routing applications in CP. In general over the years, the most watched lecture videos have been the videos on global constraints, sequence variables, reification, and routing. We can likely assume that these videos reflect the topics the students found most challenging or needed the most review on. Global constraints and reification are covered in the first half of the course, while scheduling are routing are covered in the second half. Thus, the Georgia Tech CP course does well at distributing these challenging topics.

\begin{figure}[H]
\caption{Average weekly student interaction with the course material over the semester}
    \centering
        \includegraphics[height=0.22\textwidth]{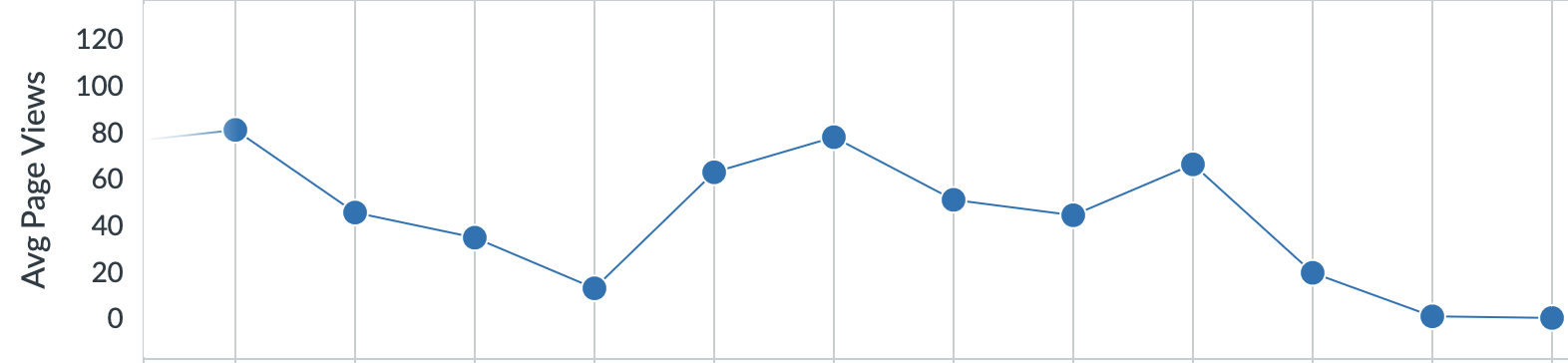}
        \label{fig:interaction}
        \end{figure}

As far as patterns of interaction within the interactive session, there have been three groups of students that have been generally observed in the course. One group of students attend every interactive session no matter what. The second group of students attend sporadically, which is usually only when they need help and at times close to assignment deadline. The last group of students function independently and don't come to any interactive sessions. The first group of students are generally strong performers in the course and understand the concepts well. The second group of students tend to be grade motivated and sometimes perform well, but do not usually build a deeper understanding of CP. The third group is a unique case where some students establish a good enough understanding from the videos and don't need the sessions while other students in the group are falling behind and struggle in the course. It is important to look at the assignments of students in this third group to identify if they need extra support or not. In addition to the grade received on the assignment, it is also helpful to look at the way constraints are written to see if the students has a certain maturity in their modeling. After all, there is a large difference between a concise and effective constraint and a hardcoded constraint specific to the dataset, even if they produce the same result.

\paragraph*{Reception}
To date, there have been six completed CP courses at Georgia Tech between Fall 2018 and Spring 2023. With each iteration of the course, the popularity of the course within the Industrial and Systems Engineering department has grown. Table~\ref{tab:enrollment} shows the number of students who took each iteration of the Georgia Tech CP course. The Spring 2023 course was intentionally set to a lower capacity as it was the first time the course was offered in back-to-back semesters. At the time of writing, the upcoming Fall 2023 course has 96 students registered with a further 33 on the waitlist. Enrollment in the Georgia Tech CP course has had full registration with spillover onto waitlists in every semester since Fall 2019.

\begin{table}[H]
\caption{Enrollment for past iterations of the Georgia Tech CP course}
\centering
\begin{tabular}{|l|l|l|l|l|l|l|} 
\hline
Semester   & Fall 2018 & Fall 2019 & Fall 2020 & Fall 2021 & Fall 2022 & Spring 2023  \\ 
\hline
Enrollment & 26        & 41        & 94        & 100       & 183       & 30           \\
\hline
\end{tabular}
\label{tab:enrollment}
\end{table}

Despite the growth in enrollment, the quality of the Georgia Tech CP course has been maintained. Surveys were given to all students in the last five iterations of the course. Some questions with average ratings on a 1-5 scale, with 1 being the lowest possible rating and 5 being the highest possible rating, are shown in Table~\ref{tab:ratings} below. The course has maintained consistently high ratings since its inception.

\begin{table}[H]
\caption{Survey ratings for past iterations of the Georgia Tech CP course}
\centering
\begin{tabular}{|l|l|l|l|l|l|} 
\hline
Semester                       & Fall 2019 & Fall 2020 & Fall 2021 & Fall 2022 & Spring 2023  \\ 
\hline
Number of Respondents          & 15        & 82        & 93        & 153       & 26           \\ 
\hline
Amount Learned                 & 4.9       & 4.8       & 4.5       & 4.4       & 4.6          \\ 
\hline
Instructor stimulates interest & 4.9       & 4.95      & 4.8       & 4.6       & 4.8          \\ 
\hline
Instructor effectiveness       & 5         & 4.97      & 4.9       & 4.7       & 4.7          \\ 
\hline
Course effectiveness           & 4.9       & 4.92      & 4.8       & 4.3       & 4.6          \\
\hline
\end{tabular}
\label{tab:ratings}
\end{table}

Some written feedback from the surveys can be seen below:

\begin{itemize}
    \item ``The structure was well organized to incrementally build our familiarity with the material. The flipped classroom format was also effective, in being able to review the recordings several times, as well as being able to revisit the recordings later in the semester. The in class sessions were useful to solidify our understanding.''
    \item ``The lecture videos were high quality and engaging. The professor themed the entire course and the amount of effort he put into making it work in a virtual format is commendable! I loved that I could watch and re-watch the lecture videos as many times as I needed to. The interactive sessions were a must have. Getting to interact with the professor after having watched the lecture videos and read through the assignments really helped and he generally gave great pointers and advice. The breakout rooms were must haves as well as often the questions I wanted to ask would not have been appropriate to ask to the class. Finally, I thought the assignments were incredible benchmarks for whether or not we fully grasped and could synthesize many concepts from the course.''
    \item ``I really enjoyed a lot of the assignments, and the theme of the course was perfect. I am a huge star wars fan (currently getting through all of the animated shows which I highly recommend), and the fun assignments made me want to complete them and I always looked forward to the story of the assignments. The interactive sessions were also super helpful and an amazing idea. I always got great help there especially in the 1-to-1 breakout rooms.''
    \item ``It is hard to say because there were so many things that I loved about the way this class was taught. 1. I would say the interactive sessions (at a convenient (5 PM) after my full-time job) and the Ed Discussion were some of the best features for getting help; I felt like I was able to get the face-to-face and virtual support I needed way more than in other classes I've taken. The professor and TAs went out of their way to make themselves available in this class, way more than in other classes. In other classes, I have had to seek out help more, and it is harder as a Distance Learning student to ask questions without the ability to do so in real-time. But this class does it! 2. As a close second, having our grades be based on assignments is the way I like to learn -- learning by applying and doing (rather than solely being tested on theory).''
\end{itemize}

In addition to the favorable enrollment and positive survey results, the Georgia Tech CP course has received further accolades. The course received the Teaching Excellence Award for Online Teaching and the Student Recognition of Excellence in Teaching: Class of 1934 Award for the Fall 2020 iteration of the course. Furthermore, the course received the Student Recognition of Excellence in Teaching:  CIOS Honor Roll for the Fall 2020, Fall 2021, and Fall 2022 iterations of the course, an award reserved for teaching excellence with large class sizes.

\section{Lessons Learned}
\label{lessons}

Over the many iterations of the Georgia Tech CP course, both the authors have gained numerous insights into the learning process. Van Hentenryck has been instructing the course since its inception, while Santanam has been involved in every iteration of the course for the last four years as either a student, teaching assistant, or instructor. Some key areas of emphasis are expounded upon below.

\paragraph*{Teaching to Undergraduate Students}

One of the unique parts of the Georgia Tech CP course is the large presence of undergraduate students each year. In every iteration of the course, undergraduate students have comprised at least 75\% of the class. Given the audience, this necessitates a different approach to CP than what one might generally see at most universities. Given their age and relative inexperience in the field, it is much more important to cultivate an interest in CP (or optimization in general) in each student. The Bachelor of Science in Industrial Engineering degree at Georgia Tech has specializations in analytics, statistics, economics, operations research, supply chain, and a general studies track as well. It is very likely that the majority of students in the class have other backgrounds and interest. Thus, many who take the CP course are doing so to learn something new, rather than trying to build upon a longstanding interest. It is for this reason that teaching in an engaging way is important. Presenting the real-world applications of CP and allowing undergraduate students to get hands-on with modeling early in the course is vital to enable a student to realize their interest or disinterest in the subject. A focus on real-world applications is also helpful for an undergraduate audience due to the large percentage of undergraduate students that go straight to work in industry upon completing their degree. For them, the focus is on developing the modeling skills and the problem-solving mindset over the theoretical underpinnings one might expect a graduate student to care more about. A last distinctive adaptation needed when teaching to undergraduate audiences is the role of coding in a CP course. At an undergraduate level, many students lack the coding maturity that allows one to take a new programming language and efficiently write the necessary syntax to execute on an ideated model. Learning a new programming language in OPL is a difficult task for many, and most undergraduate students lack the full ability to teach themselves new syntax solely by reading documentation. Undergraduate students require and continually ask for a good deal of coding examples and demonstrations, as well as help debugging and interpreting documentation. Thus, when teaching to undergraduate students, it is important to make the coding parts of a course as accessible as possible. One must decidedly think of coding exercises as a means to enable the building of CP-related intuition in undergraduate students and lower the barriers to entry there.

\paragraph*{Teaching to Engineering Students}

Another novel aspect of the Georgia Tech CP course is that it is housed within the Industrial and Systems Engineering Department inside the College of Engineering. The majority of CP courses around the world are taught by computer science faculty to computer science students, as opposed to the setup at Georgia Tech. Teaching CP to engineering students also requires careful consideration in pedagogical choices. Since the students hail from an engineering background, they tend to be very focused on the impact that CP-driven processes have on real-world systems. Most students seem to be more motivated by those outcomes CP can drive rather than by the inner workings of CP itself. For that reason, the Georgia Tech CP course has videos on real-world case studies. For example, Van Hentenryck presents two videos detailing applications of CP for integrated container terminal operations and scheduling at ports. The coding aspect of CP is also somewhere where engineering students may not be as advanced as their computer science peers. At Georgia Tech, students only have exposure to two semesters of Python before taking Constraint Programming. This can make it difficult for some students in learning a new programming language, and may lead to difficulties in debugging or expressing complex constraints.

\paragraph*{Modeling-Focused Teaching}

One of the main ways that the Georgia Tech CP course makes the subject more accessible at an introductory level is through a strong focus on writing CP models. The emphasis on modeling helps students understand how to breakdown a problem from a CP perspective and express it via a set of constraints. Since the students (particularly undergraduates in engineering) are motivated by the applications of CP, the use of black-box solvers serves to expedite the process as a whole and allows a student to go from problem to results and output much quicker. Largely limiting the instruction and assessment to modeling enables instructional staff to reclaim time that can be used to add depth or breadth in applications of CP. By way of the modeling emphasis, students also learn about elegance in model formulation and are encouraged to write constraints multiple different ways. Students end up generating a wide variety of creative formulations. Details and concepts behind search procedures and CP solvers are covered in the course, but the emphasis on the assignments largely shifts away from that.

\paragraph*{Autograders}

In order to teach the Georgia Tech CP course to hundreds of students every year, an autograder system had to be implemented for efficient grading. Manually having instructors and teaching assistants run each model and verify each solution would be inefficient. For the Georgia Tech CP course, a master Python script takes all the model files submitted by the students and runs them a few at a time. The model files produce an output in a format that is pre-specified in the assignment instructions. Students are encouraged to model however they want with whatever variables they want. Normally, a wide range of formulations are seen. The only ask is to convert their output into the pre-specified output which usually comprises a trivial post-processing step. An OPL script is then used to verify if the students' solutions are feasible and/or optimal. The results on each model's correctness is written to a CSV file that instructional staff can use for grading. Students who did not pass the autograder have their models manually checked for partial credit and are given feedback on where errors were committed. Without this kind of autograder system, countless additional hours would have to be spent by course instructional staff every semester. The lack of an autograder system would mean having to set a lower capacity limit on the course. Thus, such a system enables the spreading of CP education to larger numbers of students and reduces the marginal cost of time and effort required for each additional students. In general, any course, whether in-person or online, or MOOC or university-based, would be recommend to have an autograder. Thanks to the fact that we can verify the solutions to NP-complete problems in polynomial time, the autograder ensures fairness and consistency in grading, doing so in a timely manner.

\paragraph*{Engaging Large Audiences}

Ensuring that every students stays engaged, interested, and has their educational needs met is one of the most challenging things to do as an instructor. Especially when delivering CP MOOCs or large CP classes over 100 students, connecting with each and every students may seem like a daunting tasks. There are, however, a few strategies that one can employ. The first method is to make the learning environment and process as fun as possible. In the Georgia Tech CP course, this is achieved through immersion in the Star Wars theme. All videos and assignments are tinged with Star Wars lore, characters, locations, and more. The illusion is kept up in both verbal and written communication from the instructional staff, who only use the Star Wars names for themselves. Within the interactive sessions on Zoom, instructional staff also have their Zoom backgrounds set to Star Wars-related imagery. For many students, this can bring a lightheartedness to the class by relating a new, daunting subject with something familiar. Other students may also find the journey of ``becoming a Jedi`` motivating and place themselves within the story as they strive to complete their assignments as part of their mission. In addition to a fun theme, engaging large audiences required multiple potential points of engagement with students. In the Georgia Tech CP course, students can access the content via lecture videos 24/7, while also having interactive sessions multiple times a week. These interactive sessions allow for tighter knit review, in addition to one-on-one interactions with each student in attendance. For students with schedule or time zone issues (especially in the case of online courses and MOOCs), interaction via a discussion or Q\&A forum is yet another option. Lastly, ad hoc touchpoints with any member of the instructional staff makes learning and assignment help available for every student. The last method for large audience engagement is about accommodation of different learning styles. For those who learn by watching, videos are a great solution. Those who prefer to read can review the posted PDF lecture slide handouts. Students who learn ``by doing`` can code along with demos and gain numerous opportunities for practice through the weekly hands-on assignments. Students who learn through one-on-one discussion find the breakout rooms in the interactive sessions useful, while those students who do not need any help on particular assignments are not required to engage with any of the resources; the materials and touchpoints for help are there as and when each student needs them.

\paragraph*{Teaching via Distance Learning}

Delivering a successful online course requires a reliable set of technologies to produce a quality student experience. It can never be overstated how important good audio and video equipment is in creating high-quality videos. Students need to be able to clearly see and hear instruction in order to properly internalize it. Additional work in animation and graphics is helpful but not required; clear content is more important than theatrics. Beyond lecture video production, tools for interactive video sessions and breakout rooms are also vital for seamless transition between small group and one-on-one environments. The Georgia Tech CP course has used both Zoom and BlueJeans effectively in the past. A proper discussion forum that handles question and answer posts between students and instructional staff is a useful tool for communication, debugging, and conceptual help offline outside of interactive sessions. The Georgia Tech CP course has used both Ed Discussion and Piazza. In a large online environment where students are not all located in the same place, empowering students to help other students in the discussion forums foments greater understanding of the material between students. This can happen holistically, but it may behoove a CP instructor to offer extra credit or some similar incentive for answering the questions of peers. For overall course structure, housing course materials, announcements, and links in a learning management site like Canvas, Sakai, or Google Classroom can also be beneficial organizationally. Lastly, it is important to check in with students that one may see really struggling or falling behind. Without seeing the students physically in the classroom, it can be easy for students to fall through the cracks due to the isolation. Instructional staff may also not be fully aware of the extent of a student's personal situation. For any students that are falling behind, it is generally good practice to reach out and inquire on ways to accommodate or assist them.

\section{The Future of CP Education}
\label{sec:future}

The lessons presented above offer some thoughts and guidance for constructing a successful CP course in modern times. However, looking ahead to the future, there will need to be further adaptations and changes within the CP community as a whole to ensure the long-term growth of CP in both the education and research spheres.

\paragraph*{Promotion of CP}
One of the main obstacles facing CP education going forward has to do with promotion. The fact of the matter is, CP is a form of artificial intelligence (AI). However, this fact is grossly understated in educational spheres. The machine learning (ML) community has so tightly hitched their wagon to AI that many members of the general (and scientific) public often conflate the two and use the terms AI and ML interchangeably. This is not to say that CP must suddenly match ML in terms of media coverage and hype, but the way the CP community positions the subject can have a great impact on the desire of new students wanting to take an introductory course in CP. If one were to describe CP as AI that solves Sudoku in milliseconds and powers delivery systems within massive supply chains, this line of messaging could be more appealing to young students today. The challenge of CP relative to fields like ML is also something that could be spoken about more when promoting to young students in computer science and engineering. ML at an introductory level is largely a problem of organizing data appropriately and putting it into a black-box function that provides a trained model. On the other hand, even at the most introductory stages, CP involves writing out a uniquely customized and specific model for each problem and going a step beyond data organization in order to solve problems. The additional skill needed here is something that could prove appealing to students and motivate them in their CP journey.

\paragraph*{Introductory CP Resources}
Another positive development with regards to CP education would be greater availability of introductory resources, especially at the undergraduate level. To use the ML community as a point of comparison yet again, undergraduate learners can easily sink their teeth into \textit{An Introduction to Statistical Learning} \cite{james2013introduction}, which is also fairly standardized and used at many universities across the world. The mathematical programming community also has fairly standard undergraduate-level introductory texts like \textit{Optimization in Operations Research} \cite{rardin1998optimization} and \textit{Operations Research: Applications and Algorithms} \cite{winston2022operations}. It would be beneficial to the CP community to have such approachable texts with relative standardization in adoption across universities. The focus of this kind of text should be strongly in the applied realm. From experience, undergraduate students are often loathe to dive into textbooks with numerous theorems and proofs. Rather, they would prefer a book with numerous worked examples and code samples showing toy examples of real-world applications. A book that does this well in the ML field is \textit{Hands-on machine learning with Scikit-Learn, Keras, and TensorFlow} \cite{geron2022hands}. Of course, a textbook is not the only solution to this problem, but a standardized collection of introductory-level information along with a guide or outline that instructors can follow would be extremely beneficial, be it a textbook or an organized website.
However, in presenting application-driven textbooks and other resources, the CP community must also standardize more on the set of tools used to introduce coding to students. In the ML world, most students first use Python with libraries like Scikit-learn \cite{pedregosa2011scikit}. In the CP world, there are multiple different languages used across many introductory courses. Rather than potentially overwhelming students with the choice between OR-Tools, MiniZinc, CPLEX, Mini-CP, and more, it would be better for the community as a whole to focus in on specific languages and solvers to use when educating newcomers to the subject. It may seem unlikely after 40 years of CP that the community would suddenly align on a set of tools, but adjustments to at least focus on a smaller subset would be a step in the right direction.

\paragraph*{Availability of CP Courses}
Further proliferation of CP courses at both the university level and MOOCs is vital for the survival and growth of CP as a practice. Comparing again to ML, almost every university with a computer science, math, or engineering department these days has at least one course in ML. Similarly, there are innumerable websites with online courses for ML topics at all levels. The CP community should place special focus on outreach to schools without CP courses or faculty members with a CP background. Furthermore, continued promotion of existing MOOCs along with the development of new MOOCs will help perpetuate a virtuous cycle. Without new students and the continued development of those starting their CP journeys, new researchers in CP cannot be trained and cannot go on to spread their ideas around the world. The fact of the matter is that CP education and CP research go hand in hand. One cannot survive without the other. Thus, a focus on teaching CP will ultimately derive benefits in the research sphere as well. Additionally, the updating of centralized lists of CP courses like the one found on \href{http://www.pearltrees.com/constraints/courses/id39842792}{Pearltrees} (\href{http://www.pearltrees.com/constraints/courses/id39842792}{http://www.pearltrees.com/constraints/courses/id39842792}) would be helpful to students in finding educational resources, as well as helpful to instructors to gain inspiration on ways to improve their own CP courses.

\paragraph*{The Impact of Large Language Models}

\begin{figure}[!t]

    \centering
        \includegraphics[height=0.5\textwidth]{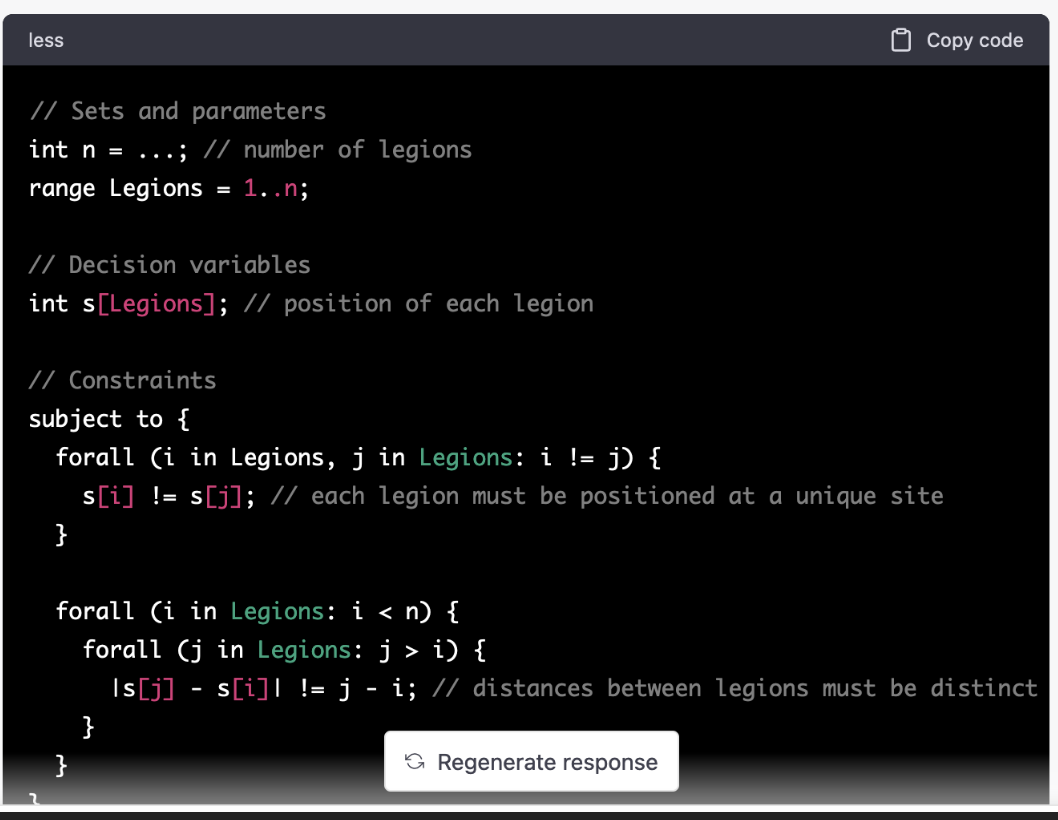}
        \caption{A Model Produced by ChatGPT for an Assignment in Resource Allocation.}
        \label{fig:chatGPT}
        \end{figure}

Large Language Models (LLMs) represent a fundamental challenge / threat for every project-based course. Constraint programming is no exception. ChatGPT was run on the assignments for the Georgia Tech CP course. The results were stunning. On the early assignments, ChatGPT essentially produced the correct models. As the assignments got more involved, ChatGPT output models that were close to the solutions, but typically had syntax or type errors. However, these models certainly give students a strong basis to start from. Figure \ref{fig:chatGPT} shows a model for an assignment that abstracts a real problem. This assignment is the first in the second part of the course, where the projects get more realistic. ChatGPT, at this point, struggles for the third part of the class, which is heavily based on scheduling and routing. 

Assignments are changed every year in the Georgia Tech CP class, but they build around the same core problems. If students have access to a model from previous years (which they are forbidden to do), it is conceivable that LLMs will fill the gap, putting increasing burden on instructors to fundamentally change assignments each year and create assignments that are radically different. Or, perhaps, in a world where LLMs will become a fundamental tool, it becomes important to rethink entirely how modeling and problem-solving courses are taught.  

\section{Conclusion}
\label{sec:conclusion}

The Georgia Tech CP course detailed in this paper demonstrates some novel ways to teach an introductory online CP course. The lecture videos and interactive sessions provide a fun way to mix large-scale instruction with focused efforts in small groups and one-on-one settings. The focus on modeling and solving applied problems helps hone the problem solving intuition of students while shoring up their coding skills.

The paper also discusses the adaptations needed when working with students from backgrounds like undergraduate engineering. Further observations are made on the use of technology in autograding and distance learning. Some thoughts from the authors on future directions of CP education (especially at an introductory level) are also discussed along with actionable recommendations for their implementation.

Ultimately, the experiences and thoughts put forward in this paper only comprise observations in a small segment of the CP community. It would be most beneficial to have discussions with CP educators around the world to learn more about other ways of effectively teaching the subject.



\bibliography{lipics-v2021-sample-article}

\end{document}